\documentstyle[eqsecnum,epsfig,psfrag,aps,floats,preprint]{revtex}

\newcommand{\be}{\begin{equation}}
\newcommand{\ee}{\end{equation}}
\newcommand{\ba}{\begin{eqnarray}}
\newcommand{\ea}{\end{eqnarray}}
\newcommand{\no}{\nonumber}
\def\laq{~\raise 0.4ex\hbox{$<$}\kern -0.8em\lower 0.62

ex\hbox{$\sim$}~}

\def\gaq{~\raise 0.4ex\hbox{$>$}\kern -0.7em\lower 0.62
ex\hbox{$\sim$}~}

\title{\begin{flushright}{\rm \small
CERN-TH/2002-064\\
gr-qc/0203093} \\
\end{flushright}
Black holes from high-energy beam--beam collisions}
\author{E. Kohlprath\footnote{Emmanuel.Kohlprath@cern.ch} and
 G. Veneziano\footnote{Gabriele.Veneziano@cern.ch}}
\address{\sl Theory Division, CERN, CH-1211 Geneva 23, Switzerland} 

\begin{document}

\maketitle

\begin{abstract}
Using a recent technique, proposed by Eardley and Giddings, we extend their
 results to
 the high-energy collision of two beams of massless particles, i.e. of two
 finite-front shock waves. Closed (marginally) trapped surfaces  can be
 determined analytically in several cases even for collisions at
 non-vanishing impact parameter  in  $D\ge 4$  space-time dimensions.
 We are able to confirm and extend earlier conjectures by Yurtsever, and 
 to deal with arbitrary axisymmetric profiles, including an
 amusing case of ``fractal'' beams. We finally discuss some
 implications of our results in high-energy experiments and in cosmology. 
\end{abstract}
\vfill
\begin{flushleft}
CERN-TH/2002-064\\
March 2002\\

\end{flushleft}

\newpage
\section{Introduction}\label{sec1}

The gravitational collapse induced, in classical General Relativity (GR),  by
 trans-Planckian-energy collisions of particles and/or waves has
 attracted much theoretical attention since the early seventies.
 In the pioneering papers by Khan and Penrose \cite{KP} and by Szekeres
 \cite{Sze} (see also  \cite{Yur1}, \cite{DT}),
 the case of infinitely-extended homogeneous plane waves
 was  solved analytically in the interaction region where a
 (naked) singularity is inevitably produced.
 At the other extreme, the scattering of point-like objects (or of black holes)
 at zero impact parameter was investigated in classic papers by D'Eath and
 Payne \cite{D'P}, while R. Penrose \cite{Penrose} managed to obtain a
 rigorous lower bound on the fraction of incident energy
 ending up in the black hole inevitably resulting from the collision.
 The difficult, intermediate case of the collision of 
 finite-front shock waves has received comparatively
 little attention, a noteworthy exception  being ref. \cite{Yur2}.

Somewhat more recently, trans-Planckian scattering of particles and strings
 were investigated at the quantum level, as a gedanken experiment aimed at
 answering some fundamental questions in quantum gravity and/or in string
 theory. The problem turned out to be tractable either at large impact
 parameters ($b \gg R_s \sim (GE)^{1/(D-3)}$) \cite{'tH},
 through an eikonal approximation,
 or when string-size effects manage to
 screen \cite{ACV} the non-linear classical effects that
 should trigger a collapse. In either case
 no black hole is formed. In spite of much effort, the region 
 $b \le R_s$, where black holes are expected
 to form, has remained untractable.

The renewed recent interest in the field stems mainly from the
following motivations:
\begin{itemize}
\item String-inspired cosmological models, such as the
 pre-big bang scenario \cite{PBB}, connect (dilaton-driven) inflation 
 to gravitational collapse through a conformal change of the metric. 
 In order to avoid fine-tuning the initial conditions, it is crucial that
 collapse occur as generically as possible. The case of spherical symmetry
 was addressed in \cite{BDV}, while that of exact planar symmetry was solved
 analytically in \cite{FKV},  using precisely the techniques of \cite{KP},
 \cite{Sze}, and  \cite{Yur1}. Since exact spherical or planar symmetry are
 quite special, it looks very desirable to extend the calculations to the case
 of finite-front shock waves, but, so far, very little progress was made, if
 any. 
\item The idea of large extra dimensions and of the brane Universe \cite{extra}
 allows for gravity to become higher-dimensional, and stronger than usual,
 below almost macroscopic distances. The true scale of quantum gravity could
 become as low as a few TeV. In such a context, black-hole formation in one of
 the near-future accelerators is all but excluded (for recent work on the
 subject, see \cite{G} and \cite{GRW} and references therein), but there has
 been some debate \cite{Volo} as to the actual value of the cross section for
 black-hole formation. 
\item New cosmological models based on the brane Universe idea have recently
 been proposed \cite{ekp}. In these models the big bang event is
 identified with the instant at which two almost parallel branes collide.
 Although the branes move slowly in this case, it is possible that techniques
 used for the relativistic case can be generalized to other situations endowed
 with similar symmetries.
\end{itemize}

Recently,  Eardley and Giddings \cite{EG} proposed a promising technique
 for determining the occurrence of closed trapped surfaces (CTSs) in the
 collision of two shock waves. They considered the case of two colliding point
 particles at generic impact parameter $b$ and in any $D \ge 4$. They succeeded
 in constructing CTSs for general $b$ in $D=4$, and at $b=0$ for $D >4$, and
 offered arguments in favour of the existence of a CTS  also  in $D>4$ for
 sufficiently small $b$. A consequence of their results is a lower limit on
 the cross section for BH formation from point-like particles. However, it
 is well known \cite{ACV} that strings behave rather as quasi-homogeneous
 beams over a size of order $\lambda_s$, the (quantum) string-length
 parameter. As we shall discuss at the end, this could reduce somewhat the
 lower bound on the cross section.

With all these motivations in mind we shall extend the work of
 \cite{EG} to the case of finite-size beam--beam collisions.
 We will first review, for completeness, the argument of \cite{EG} and then
 apply it to the case of homogeneous beams of finite size, first for $D=4$ and
 $b=0$, then for $D=4$ and $b\ne 0$ and then for $D>4$ and any $b$. Finally, we
 shall present results for axisymmetric collisions when the profile of the
 beam is arbitrary and discuss some physical implications of our results.

\section{A criterion for gravitational collapse} \label{sec2}

An impulsive wave moving with the speed of light  along the positive $z$ axis  
 of a \mbox{$D$-dimensional} space-time leads to the metric (see e.g.
 \cite{AS}, \cite{DT}, \cite{FPV})
\be
ds^2=-d\bar u\,d\bar v+\phi(\bar{\bf x})\delta(\bar u)\,d\bar u^2
+d\bar{\bf x}^2,\label{barmetric}
\ee
where $\bar u=t-z$, $\bar v=t+z$ and $\bar{\bf x}$ are the $d \equiv (D-2)$
 transverse coordinates. Einstein's equations  require
\be
\Delta \phi(\bar{\bf x})=-16\pi G\rho(\bar{\bf x}),
\ee
where $\rho(\bar{\bf x})$ is the energy density (energy per unit transverse
 hypersurface) in the beam. Let us use from now on units in which $8\pi G =1$
 and $\rho$ has (for any $D$) dimensions of an inverse length. 

In the coordinates
 $\bar u,\bar v,\bar{\bf x}$ the geodesics and their tangent vectors are
 discontinuous across the shock \cite{DT}, \cite{FPV}, \cite{EG}, whereas in
 the new coordinates
\ba
\bar u&=&u\\
\bar v&=&v+\phi({\bf x})\theta(u)+\frac{1}{4}u\theta(u)
(\nabla\phi({\bf x}))^2\\
\bar x^i&=&x^i+\frac{u}{2}\theta(u)\nabla_i\phi({\bf x})
\ea
they are continuous. In these coordinates the metric becomes \cite{EG}
\be
ds^2=-du\,dv+H_{ik}H_{jk}\,dx^i\,dx^j,
\label{wave1}
\ee
where
\be
H_{ij}=\delta_{ij}+\frac{1}{2}\nabla_i\nabla_j\phi({\bf x})u\theta(u).
\ee

Let us now consider the collision of two particle beams, or shock waves,
 moving in opposite directions along the $z$ axis. By causality, outside  the
 interaction region $u>0,v>0$, the metric is  given  by a trivial superposition
 of two metrics of the form (\ref{wave1}):
\be
ds^2=-du\,dv+\Bigl[H_{ik}^{(1)}H_{jk}^{(1)}+H_{ik}^{(2)}H_{jk}^{(2)}
-\delta_{ij}\Bigr]\,dx^i\,dx^j,\label{metric}\label{wave12}
\ee
where
\ba
H_{ij}^{(1)}&=&\delta_{ij}+\frac{1}{2}\nabla_i\nabla_j\phi_1({\bf x})
u\theta(u)\\
H_{ij}^{(2)}&=&\delta_{ij}+\frac{1}{2}\nabla_i\nabla_j\phi_2({\bf x})
v\theta(v)\\
\Delta \phi_{1,2}({\bf x})&=&- 2 \rho_{1,2}({\bf x}). 
\ea

We reproduce now, for completeness, the construction of \cite{EG} to find
 a (marginally) closed trapped surface (CTS) ${\cal S}$ lying in the union
 of the two null hypersurfaces $u=0,v\leq 0$ and $v=0,u\leq 0$. This property
 of ${\cal S}$ allows us to use the simple block-diagonal metric
 (\ref{metric}). On the  null hypersurface $u=0,v\leq 0$, the
 non-vanishing components of the Christoffel connection are simply
\be
\Gamma^v_{\ ij} = 2 \Gamma^i_{\ uj} = \nabla_i\nabla_j
\phi_1({\bf x})\label{Gammas}.
\ee
Actually, this result is strictly valid for $u > \epsilon >0$ with 
$\epsilon$ arbitrarily small, so that we resolve a possible ambiguity
by defining $\theta(0) = \theta(\epsilon) =1$.
Similarly, on the  null hypersurface $v=0,u\leq 0$, the
 non-vanishing components of the Christoffel connection are
\be
\Gamma^u_{\ ij} =  2 \Gamma^i_{\ vj} = \nabla_i\nabla_j
\phi_2({\bf x}).
\ee

Let us define ${\cal S}={\cal S}_1\cup{\cal S}_2$, with
 ${\cal S}_1$: $u=0,v=-\psi_1({\bf x})\leq 0$, and ${\cal S}_2$: $v=0,
 u=-\psi_2({\bf x})\leq 0$; then ${\cal S}$ intersects the $d$-dimensional
 hypersurface $u=0=v$ on a closed $(d-1)$-dimensional hypersurface ${\cal C}$.
 We recall \cite{HawkingEllis} that a CTS  is a $C^2$ closed space-like
 $d$-dimensional (hyper)surface ${\cal S}$ such that the two families of null
 geodesics orthogonal to ${\cal S}$ are converging at ${\cal S}$. Modulo other
 conditions on the energy-momentum tensor that are met in our case, the
 existence of a CTS guarantees the occurrence of gravitational collapse,
 i.e., typically, the emergence of singularities hidden behind black-hole
 horizons. Rather than for CTSs we will look for  marginally trapped
 (hyper)surfaces  (MCTSs), on which the above null geodesics have zero
 convergence.
In general ${\cal S}$ will be defined by $f_1=0,f_2=0$, where $f_1$ and $f_2$
 are $C^2$-functions such that $f_{1;\mu}$ and $f_{2;\nu}$ are non-vanishing,
 non-parallel, and satisfy
\be
(f_{1;\mu}+\mu f_{2;\mu}) \ (f_{1;\nu}+\mu f_{2;\nu})g^{\mu\nu}=0
\ee
for two distinct real values $\mu_1$ and $\mu_2$ of $\mu$. Let $N_1^\mu$ and
 $N_2^\mu$ be two null vectors normal to ${\cal S}$ and proportional to
 $g^{\mu\nu}(f_{1;\nu}+\mu_1 f_{2;\nu})$ and $g^{\mu\nu}(f_{1;\nu}+\mu_2
 f_{2;\nu})$, normalized by $N_1^\mu N_2^\nu g_{\mu\nu}=-1$ and let $Y_a^\mu$
 ($a = 1, 2, \dots, d$) be a set of space-like
 unit vectors orthogonal to each other and to
 $N_1^\mu$ and $N_2^\mu$. The two null second fundamental forms of
 ${\cal S}$ are defined by \cite{HawkingEllis}
\be
\chi_{n\mu\nu}=-N_{n\rho;\sigma}(\sum_a Y_a^\rho Y_{a\mu})~ 
(\sum_b Y_b^\sigma Y_{b\nu}).
\ee
${\cal S}$ is a CTS (MCTS) if $g^{\mu\nu}\chi_{1\mu\nu}$ and
 $g^{\mu\nu}\chi_{2\mu\nu}$ are never positive (vanish) on ${\cal S}$. 
On ${\cal S}_1$ we choose
\be
N_1^\mu=\left(\begin{array}{c}0\\-1\\0\\0\\\dots\\0
\end{array}\right), N_2^\mu=\left(\begin{array}{c}-2\\-\frac{(\nabla\psi_1)^2}
{2}\\ \psi_{1,1}\\\psi_{1,2}\\\dots\\\psi_{1,d}\end{array}\right),
 Y_1^\mu=\left(\begin{array}{c}0\\-\psi_{1,1}\\1\\0\\\dots\\0\end{array}
\right), Y_2^\mu=\left(\begin{array}{c}0\\-\psi_{1,2}\\0\\1\\\dots\\0
\end{array}\right) , \dots 
\ee
One then easily finds that
\be
g^{\mu\nu}\chi_{1\mu\nu} = 0, \quad g^{\mu\nu}\chi_{2\mu\nu} = \Delta (\phi_1
-\psi_1).
\ee
On ${\cal S}_2$ we choose, analogously,
\be
N_1^\mu=\left(\begin{array}{c}-1\\0\\0\\0\\\dots\\0
\end{array}\right), N_2^\mu=\left(\begin{array}{c}-\frac{(\nabla\psi_2)^2}{2}
\\-2\\ \psi_{2,1}\\\psi_{2,2}\\\dots\\\psi_{2,d} \end{array}\right),
 Y_1^\mu=\left(\begin{array}{c}-\psi_{2,1}\\0\\1\\0\\\dots\\0\end{array}
\right), Y_2^\mu=\left(\begin{array}{c}-\psi_{2,2}\\0\\0\\1\\\dots\\0
\end{array}\right) , \dots
\ee
giving
\be
g^{\mu\nu}\chi_{1\mu\nu} = 0, \quad g^{\mu\nu}\chi_{2\mu\nu} = \Delta (\phi_2
-\psi_2).
\ee
Continuity of the outer null normal $N_2^\mu$ on ${\cal C}$ ($u=v=0$) requires
\be
\frac{\psi_{1,1}}{\psi_{2,1}}=\frac{\psi_{1,2}}{\psi_{2,2}} = \dots
\ee
i.e. $\nabla\psi_1$ to be parallel to $\nabla\psi_2$ and
\be
(\nabla\psi_1)^2(\nabla\psi_2)^2=16.
\ee
The necessary and sufficient condition for this to happen is \cite{EG}
\be
\nabla\psi_1 \cdot \nabla\psi_2= 4.
\ee
To summarize,  ${\cal S}$ is a MCTS under the following conditions:
\ba
\psi_{1,2}>0\ {\rm inside}\ {\cal C}, \psi_{1,2}&=&0\ {\rm on}\ {\cal C}
\label{condition1}\\
\Delta(\psi_{1,2}-\phi_{1,2})&=&0\ {\rm inside}\ {\cal C}\label{condition3}\\
\nabla\psi_1\cdot\nabla\psi_2&=&4\ {\rm on}\ {\cal C}.\label{condition5}
\ea
In \cite{EG} the authors construct ${\cal C}$ for the collision of massless
 point particles at zero impact parameter $b$ and any $D$, as well as
 for non-vanishing impact parameter in $D=4$. In the following sections we
 will describe extensions to the case of non-point-like beams.

\section{Homogeneous finite-size beams} \label{sec3}

Let us consider the case of finite-size beams with radius $R_1\geq R_2$ and
 homogeneous energy density $\rho_{1,2}$ inside. It is useful in this case
to introduce for each beam its ``focal distance''
\be
f_{1,2} = \frac{d}{2} \rho_{1,2}^{-1}\, , ~ (8\pi G =1).
\ee
This is where the null geodesics parallel to the $z$-axis converge after
 hitting the shock wave, which therefore acts as a perfect anastigmatic lens
 \cite{GV}, \cite{FPV}.

\subsection{$D=4$, $b=0$}

For $b=0$, $D=4 \ (d=2)$ we choose
\ba
\label{psifinitesizebeams4}
2 f_{1,2} \psi_{1,2}(r)&=& (R_{1,2}^2-r^2) \theta(R_{1,2}-r)
- 2 R_{1,2}^2 \log\left(\frac{r}{R_{1,2}}\right) ~\theta(r-R_{1,2})+c_{1,2}\\
\Delta\psi_{1,2}(r)&=&-2\rho_{1,2}\theta(R_{1,2}-r),
\ea
where $r$ is the radial coordinate in the transverse ${\bf x}$ space. The
 conditions (\ref{condition3})  are already satisfied while (\ref{condition5})
 reads
\ba
4&=&\frac{d}{dr}\psi_1(r)\frac{d}{dr}\psi_2(r)={r^2 \over f_1 f_2}
\theta(R_2-r)+ {R_2^2 \over f_1 f_2 }\theta(R_1-r)\theta(r-R_2)
+\frac{R_1^2R_2^2}{f_1 f_2 r^2}\theta(r-R_1).
\ea
If
\be
R_2 > 2\sqrt{f_1 f_2}\, ,\label{fsmallerR}
\ee
(\ref{condition5}) has two solutions:
\ba
r =  r_{c1} &=& 2\sqrt{f_1 f_2}< R_2 ~,~\label{rc1} \\
r =  r_{c2} &=& \frac {R_1 R_2}{2\sqrt{f_1 f_2}} > R_1. \label{rc2}
\ea
The first lies inside both beams, while the second is external to both. The
 physical meaning of this result is quite clear: if $R_2 > 2\sqrt{f_1 f_2}$
 there are CTSs that intersect the collision plane at any value of $r$ between
 $r_{c1}$ and $r_{c2}$ (the CTS becomes a MCTS at $r = r_{c1}, r_{c2}$). 
 An example is the surface ${\cal S}={\cal S}_1\cup{\cal S}_2$, with
\be
{\cal S}_1: u = v + (r_c^2 - r^2)/r_c = 0   \, , ~ 2f_1 < r < R_1,
\ee
so that  $g^{\mu\nu}\chi_{2\mu\nu} = 4(r_c^{-1} - (2f_1)^{-1}) <0$ and
 similarly for ${\cal S}_2$. Thus ${\cal S}$ is a CTS for
 $2\, {\rm max}(f_1,f_2)<r_c<R_2$.

In the  case
\be
R_2= 2\sqrt{f_1 f_2}\, ,\label{fsmallerRa}
\ee 
we have MCTSs intersecting the collision plane for any $r$ in $R_1>r>R_2$.
 The conditions (\ref{condition1}) finally determine the constants $c_{1,2}$.
 So we have shown that in the collision of finite-size beams with vanishing
 impact parameter in 4 dimensions a CTS forms if $R_2 \ge 2\sqrt{f_1 f_2}$.
 This result is in full agreement with one of the conjectures by Yurtsever
 \cite{Yur2}. In fact his criterion for collapse, $R_2 \gg 2\sqrt{f_1 f_2}$,
 is simply replaced by $R_2 \ge c \sqrt{f_1 f_2}$ with $c \le 2$.

\subsection{$D=4$, $b\not=0$}

For the case of non-vanishing impact parameter, a closed trapped surface can be
 constucted starting with the solution at $b=0$.
 Let the first beam be centred around $-a\leq 0$ on the $x$-axis and
 the second beam around $a$ on the $x$-axis, so that the impact parameter is
 $b=2a$.

Let us construct a solution starting with (\ref{rc1}). If the impact parameter
 fulfils the condition
\be
b \leq 2 R_2 - 4 \sqrt{f_1 f_2},
\label{conditionona1}
\ee
then on the circle
\be
{\cal C}: x^2+y^2= 4 f_1 f_2 
\ee
we are still inside both beams.
This time (unlike in the case $b=0$) we can allow in $\psi_{1,2}- \phi_{1,2}$ 
not only a constant but also a term linear in $x$. We thus choose
\ba
2f_{1,2} \psi_{1,2}(x,y)&=&-((x\pm a)^2+y^2-R_{1,2}^2) \theta(R_{1,2}^2
-(x\pm a)^2-y^2)\no\\&&- R_{1,2}^2 \log\frac{(x\pm a)^2+y^2}
{R_{1,2}^2}\theta((x\pm a)^2+y^2-R_{1,2}^2)\no\\&&+\left[
4 f_1 f_2 -R_{1,2}^2\pm 2ax+a^2\right] \, ,
\ea
so that
\ba
\Delta \psi_{1,2}(x,y)&=&-2\rho_{1,2}\theta(R_{1,2}^2-(x\pm a)^2-y^2)\\
\psi_{1,2}(x,y)&=&0\ {\rm on}\ {\cal C}\\
\nabla\psi_{1}\cdot\nabla\psi_{2}&=&4\ {\rm on}\ {\cal C}.
\ea
We conclude  that if (\ref{fsmallerR}) and (\ref{conditionona1}) hold,
 a black hole will form in the collision, the maximal impact parameter being
\be
b_{max,1}=2R_2-4\sqrt{f_1f_2}.
\ee

A similar method can be used to generalize (\ref{rc2}). Let us assume,
 for simplicity, that the two  beams are identical: $R_1=R_2=R$, 
 $\rho_1=\rho_2=\rho$. If the impact parameter fulfils
\be
0<b\leq  \frac{R^2}{ f}-2 R \label{conditionona2},
\ee
then the circle
\be
{\cal C}: x^2+y^2= \frac{R^4}{4f^2} 
\ee
is external to both beams.
Choosing
\ba
2 f \psi_{1,2}(x,y)&=&-((x\pm a)^2+y^2-R^2)\theta(R^2-(x\pm a)^2-y^2)
\no\\&&- R^2\log\frac{(x\pm a)^2+y^2}{R^2}\theta((x\pm a)^2+y^2-R^2)
\no\\&&+  R^2\log\left[\Bigl(\Bigl(\frac{ax}{r_{c2}^2}\pm
 1\Bigr)^2+\frac{a^2y^2}{r_{c2}^4}\Bigl)\frac{r_{c2}^2}{R^2}\right],
\ea
we find
\ba
\Delta \psi_{1,2}(x,y)&=&-2 \rho\, \theta(R^2-(x\pm a)^2-y^2)+ \rho
\pi R^2\delta\Bigl(x\pm\frac{r_{c2}^2}{a}\Bigr)\delta(y)
\label{extracharge}\\
\psi_{1,2}(x,y)&=&0\ {\rm on}\ {\cal C}\\
\nabla\psi_{1}\cdot\nabla\psi_{2}&=&4\frac{r_{c2}^2 \Bigl(r_{c2}^2- a^2
\Bigr)}{\Bigl(r_{c2}^2+a^2\Bigr)^2 - 4 a^2 x^2} \ {\rm on}\ {\cal C}.
\ea
The extra sources on the $x$-axes at $\pm\frac{r_{c2}^2}{a}$ always lie outside
 ${\cal C}$. The situation is now the same as for the collision of two
 particles described in \cite{EG}. We use complex variables
 $z=\frac{x+iy}{r_{c2}}$ and make a conformal transformation
\be
z'(z)=\frac{1-\left(\frac{a}{r_{c2}}\right)^2}{2\frac{a}{r_{c2}}}\log\frac{1
+\frac{a}{r_{c2}}z}{1-\frac{a}{r_{c2}}z}.
\ee
The circle ${\cal C}$ is then transformed to a new curve ${\cal C}'$
 and, in terms of the new coordinates,
\ba
\Delta' \psi_{1,2}(x',y')&=&-2 \rho\, \theta(R^2-(x'\pm a)^2-y'^2)\
 {\rm inside}\ {\cal C}'\\
\psi_{1,2}(x',y')&=&0\ {\rm on}\ {\cal C}'\\
\nabla'\psi_{1}\cdot\nabla'\psi_{2}&=&4\ {\rm on}\ {\cal C}'.
\ea
The points $(\pm a,0)$ are transformed to $(\pm a',0)=(\pm\frac{1}{2}r_{c2}
 h(a/r_{c2}),0)$ where 
\be
h(w)=\frac{1-w^2}{w}\log\frac{1+w^2}{1-w^2}
\ee
has a maximum at $w_{max}\approx 0.62519$ with $h(w_{max})\approx 0.80474$.
 Taking (\ref{conditionona2}) into account we have that this solution is
 possible for
\be
b'=2a' \le b'_{max,2}=\left\{{r_{c2}\, h(w_{max})\ {\rm for}\ R<(1-w_{max})\,
 r_{c2}\atop r_{c2}\, h(1-R/r_{c2})\ {\rm for}\ R>(1-w_{max})\, r_{c2}  }
\right. .
\ee
Note that (\ref{fsmallerR}) only fixes $R<r_{c2}$ and that
 $b_{max,2}\geq b_{max,1}$.

\subsection{$D>4$, $b=0$}

Let us generalize the results to arbitrary dimensions $D>4$. For vanishing
 impact parameter we find (recalling the definition 
 $f = (d/2) (8 \pi G \rho)^{-1}$):
\ba
2f_{1,2} \psi_{1,2}(r)&=& -(r^2-R_{1,2}^2)\theta(R_{1,2}-r)+2 R_{1,2}^{D-2}
\frac{r^{4-D}-R_{1,2}^{4-D}}{D-4}\theta(r-R_{1,2})+c_{1,2}
\label{psifinitesizebeamsD}\\
\Delta\psi_{1,2}(r)&=&- 2 \rho_{1,2}\theta(R_{1,2}-r).
\ea
The condition (\ref{condition5}) reads
\ba
4 f_1 f_2 &=& \frac{d}{dr}\psi_1(r)\frac{d}{dr}\psi_2(r) f_1 f_2 
= r^2 \theta(R_2-r)+ r^{4-D}R_2^{D-2}\theta(R_1-r)\theta(r-R_2)
\no\\&+& r^{6-2D}R_1^{D-2}R_2^{D-2}\theta(r-R_1).
\ea
Keeping in mind that, by convention, $R_2\le R_1$, let us introduce two
 length scales:
\be
f = \sqrt{ f_1 f_2} \, , ~ F = f \left(\frac{R_1}{R_2}\right)^{d/2-1} \ge f.
\ee
We then distinguish various cases according to the value of $R_2$.
\begin{itemize}
\item $R_2 < 2 f$

In this case we cannot find any MCTSs.

\item  $2 f <R_2 < 2 F$

We find two MCTSs: the first intersects the collision hyperplane at
\be
r = r_{c1}= 2 f < R_2 \label{rc1D},
\ee
i.e. inside the two beams. The second intersects at
\be
r = r_{c2}= \left(\frac{R_2^d}{4f^2}\right)^{\frac{1}{d-2}}~,~ \label{rc2D}
\ee
i.e. inside beam 1 and outside beam 2.
\item $R_2 > 2 F$ 

We find again two MCTSs:  the one corresponding to the circle of radius
 $r_{c1}$ and a second with
\be
r = r_{c3}= \left(\frac{(R_1R_2)^{d/2}}{2f}\right)^{\frac{1}{d-1}} > R_1,
 \label{rc3D}
\ee
i.e. outside both beams.
\end{itemize}
Finally, the constants $c_{1,2}$ can be easily determined from
 (\ref{condition1}).

We have thus shown that in the collision of finite-size beams at vanishing
 impact parameter in $(D>4)$ dimensions, a closed trapped surface  will form
 whenever $R_2>2f$. The radii $r_{c1}$ and $r_{c2,3}$ all correspond to MCTS
 and, by continuity, there should be genuine CTS intersecting the collision
 hyperplane at $r_{c1}<r<r_{c2,3}$.

\begin{figure}
\begin{center}
\begin{psfrags}
\epsfxsize=10cm\leavevmode\epsfbox{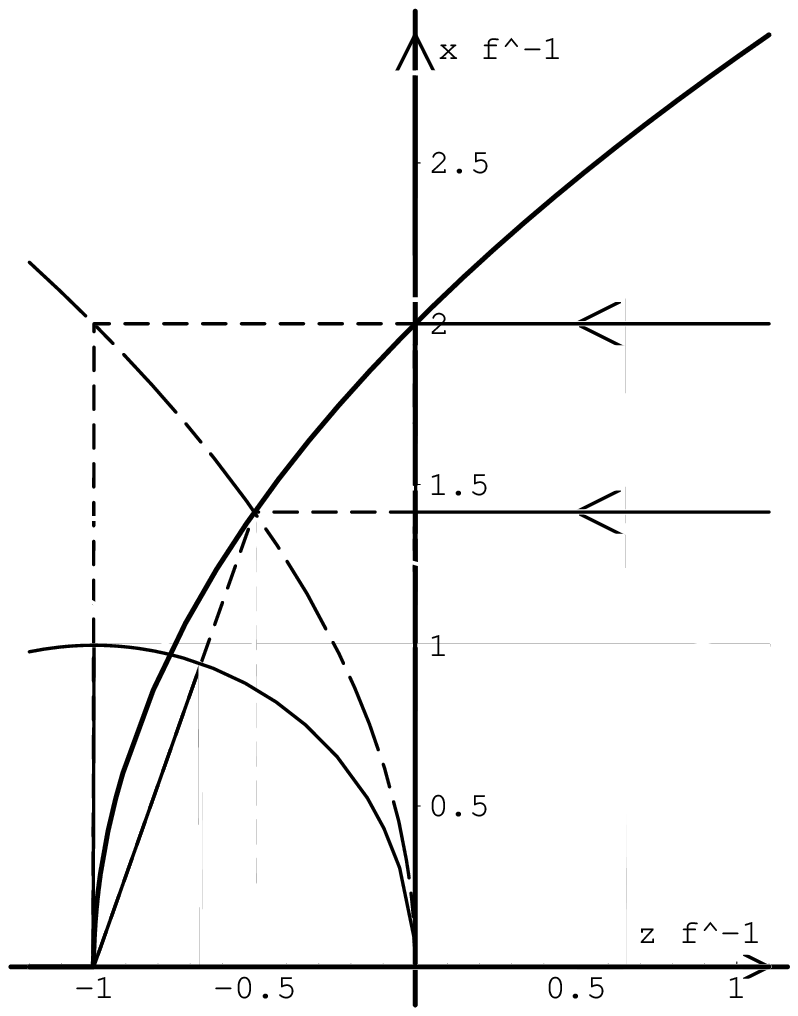}
\end{psfrags}
\caption{Null geodesics in the metric (\ref{barmetric}) for homogeneous
 beams with $f_1=f_2=f=\frac{R}{2}$ (from Ref.\protect\cite{FPV}) and the
 hypersurface $S_1\subset S$.
 The geodesics come in from the left, parallel to the $z$-axes, then jump
 according to the dotted lines and reappear on the circle $(\frac{z}{f}+1)^2
+(\frac{x}{f})^2=1$.
 All the geodesics hitting the wave at $t=z=0$ converge in $-f$ at the same
 time, the outermost coming in at 90 degrees. The bold line is the
 hypersurface $S_1:\ z=t=\frac{x^2}{4f}-f$. The dashed-dotted line is the
 hypersurface $z=-\frac{x^2}{4f}$.}
\label{figure1}
\end{center}
\end{figure}

It is amusing to try to understand, at any $D$, the physical meaning of the
critical radius $r_{c1}$ which lies inside both beams. Boost the system to
 a Lorentz frame in which $\rho_1=\rho_2=\rho$ and go back for a moment to
 the coordinates $\bar u$, $\bar v$ and $\bar{\bf x}$ to analyse the null
 geodesics. A null geodesics  parallel to the $z$-axis and initially at
 $u<0$, $v = 0$ and $R>x_0>0$ will hit the shock wave at $u=0$, and then
 instantaneously jump by \cite{FPV}
\be
\Delta t=\Delta z=\frac{\Delta v}{2}=- \frac{x_0^2}{4f},
\ee
and again follow a straight line to reach $x=0$ at 
\be
v_F=0,\quad t_F=-z_F=\frac{u_F}{2}= f.
\ee
Precisely for $x_0=2f$ we have $\Delta z=z_F$ and
 therefore a scattering angle of $\pi/2$ in the \hbox{($x$--$z$)-plane}.
In other words, for $R_2 = 2f$ the lens effect of the wave is strong enough
to deflect  the energy impinging on its edges by $90$ degrees!
 Fig.\ref{figure1} illustrates this point.

\subsection{$D>4$, $b\not=0$}

For non-vanishing impact parameter, the solution starting with (\ref{rc1D}) is
 constructed as in 4 dimensions.  If the impact parameter fulfils
\be
b\leq 2 R_2- 4 \sqrt{f_1 f_2},
\label{conditionona1D}
\ee
then the circle
\be
{\cal C}: x_1^2+\cdots+x_n^2= 4 f_1 f_2 
\ee
lies inside both beams and we can take
\ba
&&2 f_{1,2} \psi_{1,2}(x_1,\cdots,x_n)=-((x_1\pm a)^2+x_2^2+\cdots+x_n^2
-R_{1,2}^2)\no\\&&\times\theta(R_{1,2}^2-(x_1\pm a)^2-x_2^2
-\cdots-x_n^2)+\theta((x_1\pm a)^2+x_2^2+\cdots+x_n^2-R_{1,2}^2)\no\\&&
\times \frac{2R_{1,2}^{D-2}}{D-4}\left[\Bigl((x_1\pm a)^2+x_2^2+\cdots+x_n^2
\Bigr)^{\frac{4-D}{2}}-R_{1,2}^{4-D}\right]+
\left[ 4 f_1 f_2 -R_{1,2}^2\pm 2ax_1+a^2\right]~,
\ea
in order to satisfy all the conditions for a MCTS.

Therefore if $R_2>2f$ and (\ref{conditionona1D}) hold, a black hole
 will be created in the collision. To find a solution starting with
 (\ref{rc2D}) is more difficult, but it is easy to argue, by continuity, that
 an outer MCTS should exist also in this case.

\section{Non-homogeneous beams} \label{sec4}

We now want to address the question of how natural the creation of a black hole
 is in the collision of two particle beams of arbitrary
transverse profile. We will restrict our attention to axisymmetric
 examples,  i.e. to collisions at $b=0$ of beams whose
 energy density is only a function
 of the transverse radius $\rho_{1,2}({\bf x})=\rho_{1,2}(r)$. 
Then the beam energy inside a
 radius $r$ is simply
\be
E(r)=\Omega_{D-2}\int\limits_0^r\!\!dr'\,r^{'D-3}\rho(r'),
\ee
where $\Omega_d=2\pi^{d/2}/\Gamma(d/2)$ is the solid angle in d dimensions.
 We want to construct a closed trapped surface and for this we start with
\be
\psi_1(r)=\phi_1(r)+c_1,\quad \psi_2(r)=\phi_2(r)+c_2,
\ee
so that (\ref{condition3}) is fulfilled. This also implies 
\be
\frac{d}{dr}\psi_{1,2}(r)=-\frac{2 E_{1,2}(r)}{\Omega_{D-2}r^{D-3}},
\ee
so that (\ref{condition5}) can be written as
\be
1=\frac{1}{\Omega_{D-2}r_c^{D-3}}\sqrt{E_1(r_c)E_2(r_c)}.
\label{condition5alt}
\ee
If this has at least one solution, we can adjust the constants $c_{1,2}$ in
 such a way that $\psi_{1,2}$ vanishes on the critical radius, i.e. we have
 constructed a MCTS.

As a first example, consider a ``fractal'' beam
\be
\rho_{1,2}(r)=\alpha_{1,2}\ r^{\delta - d},\delta>0,
\ee
where $\alpha_{1,2}$ are constants and the last condition guarantees
that $E(r) \sim r^{\delta} < \infty$ for finite $r$. We can call $\delta$
the fractal dimension of the (energy stored in the) beam.
For $\delta \not=d-1$ we always find a critical
 radius for a MCTS at
\be
r_c=\Bigl[\frac{\delta^2}{\sqrt{\alpha_1\alpha_2}}\Bigr]^
{\frac{1}{\delta - d +1}}
\label{fracrc}
\ee
while, for $\delta=d-1$, we can choose any radius if $\alpha_{1,2}$ fulfil
the condition
\be
\sqrt{\alpha_1\alpha_2}=(d-1)^2,
\label{d-1case}
\ee
but we find no solution otherwise. We interpret this by saying that, for
 $\delta > d-1$ ($\delta < d-1$) we have  CTSs with $r > r_c$ ($r < r_c$), with
 $r_c$ given by (\ref{fracrc}), while, for $\delta = d-1$, we either have, at
 all $r$, a MCTS  (if  (\ref{d-1case}) holds), a CTS (if $\sqrt{\alpha_1
 \alpha_2}>(d-1)^2$) or, finally, nothing at all, if $\sqrt{\alpha_1\alpha_2}
 <(d-1)^2$. 

Consider finally the case
\be
\rho(r)=\alpha_{1,2}\ r^{2-d} e^{-\delta_{1,2}r^2},
\ee
where (\ref{condition5alt}) becomes
\be
2=\frac{1}{r_c^{d-1}}\sqrt{\frac{\alpha_1\alpha_2}{\delta_1\delta_2}}
\sqrt{(1-e^{-\delta_1\ r_c^2})(1-e^{-\delta_2\ r_c^2})},
\ee
and the existence of a solution depends on the precise values 
of $\alpha_{1,2}$, $\delta_{1,2}$ and
 $D$. As an example, for $D=5$, $\delta_1=\delta_2=\delta$,
 $\alpha_1=\alpha_2=\alpha= 2\delta$ there is a solution $r_c>0$ for
 $\delta>1$ but none otherwise.\\

\section{Physical implications}  \label{sec5}
Let us briefly discuss the possible physical implications of our results, first
in string cosmology and then on cross-sections for black-hole formation.
\vspace*{-.1cm}
\begin{itemize}
\item String cosmology

\vspace*{-.1cm}
We believe that our results  clearly go in favour of
the arguments given in \cite{BDV}, \cite{FKV} for the generic occurrence of
gravitational collapse in the presence of an initial, classical
 chaotic sea of massless waves. It is also self-evident
 that the criteria for collapse
never involve the Planck length/mass,
 but only dimensionless ratios of classical lengths  describing either
the geometry of the beams or the geometry of space-time. We
have thus provided further answers to the allegations of fine-tuning in
pre-big bang cosmology claimed in \cite{TW}. One limitation of our method is
 the restriction to impulsive waves. We think that this should not be a
 problem of principle and that generalization of our results to ``thick''
 waves should be possible.
By contrast, our results have nothing to say on the nature of the singularity
that lies inside the CTSs and, in particular, on whether, in its vicinity,
 space-time is described by a Kasner-like metric or by the more generic
BKL oscillatory behaviour recently discussed by Damour and Henneaux
\cite{DH}. This may very well depend on the nature of the collapsing waves.
\vspace*{-.1cm}
\item Black-hole production at accelerators

\vspace*{-.1cm}
Our results confirm the absence of the exponential suppression claimed in
 \cite{Volo} in agreement with the original estimates in \cite{G} and more
 recent work \cite{S} and \cite{H}. However they imply a revision,
 unfortunately downwards, of
 previous estimates \cite{G} and \cite{EG} of black hole production in
 theories with large extra dimensions and low-scale quantum-gravity.
 Those theories make sense only in a superstring context, which introduces
 a length scale $\lambda_s$, which is at least as large as the (true,
 multidimensional) Planck length
 $\l_P \sim M_P^{-1}$. As  mentioned in the introduction, if strings, rather
 than point particles, are colliding,  string-size effects can be modelled
 \cite{ACV} by considering beam--beam collisions with beam sizes of order
 $\lambda_s$. Using our results (\ref{conditionona1}),
 (\ref{conditionona1D}), and inserting $R \sim \lambda_s$, we find that
 the range of impact parameter where black hole formation must occur is
\be
|b| < O(R_s - \lambda_s) \, , \  R_s \sim (GE)^{1/(D-3)},
\label{newblimit}
\ee
where $E$ is the centre-of-mass energy of the collision. Equation
 (\ref{newblimit}), together with the expected relation \cite{GRW}
 $\sigma_{{\rm BH}} \sim \pi b^2$, implies that, in order to arrive at
 reasonable cross sections for black hole production, the c.m. energy should
 satisfy $E > E_{{\rm threshold}} \sim M_P (\lambda_s/l_P)^{D-3}$, i.e.
 should be parametrically larger than $M_P$ if, as expected,
 $\lambda_s/l_P > 1$. This is in agreement with many of the conclusions
 reached in \cite{GRW} (see also \cite{G1}).

\end{itemize}

\end{document}